\documentclass[aps,prd,showpacs,superscriptaddress,letterpaper,preprintnumbers,asmmath,amssymb]{revtex4}
\usepackage{amssymb,amsmath,amsbsy}
\usepackage{mathrsfs}
\usepackage[dvips]{graphics}
\usepackage{epsfig}

\def\be{\begin{eqnarray}}
\def\ee{\end{eqnarray}}
\def\bea{\begin{eqnarray*}}
\def\eea{\end{eqnarray*}}

\def\centeron#1#2{{\setbox0=\hbox{#1}\setbox1=\hbox{#2}\ifdim
\wd1>\wd0\kern.5\wd1\kern-.5\wd0\fi
\copy0\kern-.5\wd0\kern-.5\wd1\copy1\ifdim\wd0>\wd1
\kern.5\wd0\kern-.5\wd1\fi}}
\def\ltap{\;\centeron{\raise.35ex\hbox{$<$}}{\lower.65ex\hbox{$\sim$}}\;}
\def\gtap{\;\centeron{\raise.35ex\hbox{$>$}}{\lower.65ex\hbox{$\sim$}}\;}

\newcommand{\newc}{\newcommand}
\newc{\qbar}{{\overline q}}
\newc{\Kahler}{K\"ahler }
\newc{\deltaGS}{\delta_{\rm GS}}

\begin{document}
\preprint{
\vbox{\vspace*{2cm}
      \hbox{UCI-TR-2011-23}
      \hbox{October, 2011}
}}
\vspace*{3cm}

\title{Higgs Search Constraints on Fourth Generation Scenarios with General Lepton Sectors}
\author{Linda M. Carpenter}

\affiliation{Department of Physics and Astronomy  \\
   University of California, Irvine, CA 92697 \\
  lcarpent@uci.edu \\
\vspace{1cm}}

\begin{abstract}

I present a general exclusion bound for the Higgs in fourth generation scenarios with a general lepton sector. Recent Higgs searches in fourth generation scenarios rule out the entire Higgs mass region between 120 and 600 GeV. That such a large range of Higgs masses are excluded is due to the presence of extra heavy flavors of quarks, which substantially increase Higgs production from gluon fusion over the Standard Model rate.  However, if heavy fourth generation neutrinos are less than half of the Higgs mass, they can dominate the Higgs decay branching fraction, overtaking the standard $h\rightarrow WW^{*}$ decay rate.  The Higgs mass exclusion in a fourth generation scenario is shown most generally to be 155-600 GeV, and is highly dependent on the fourth generation neutrino mixing parameter. 

\end{abstract}

\maketitle

\section{Introduction}

A fourth generation is a simple and compelling extension of the Standard Model.  The most general fourth generation model consists of the addition of up and down type quarks, a charged lepton and a neutrino which most generally may have both a Dirac and  Majorana mass term.  Bounds on the fourth generation lepton sector were set by LEP II, the leptons may be quite light. Bounds on charged leptons are around 100 GeV. Bounds on unstable pure Majorana  neutrinos which decay to a W boson and a hard lepton, are 90.7, 89.5 and 80.5 GeV respectively for e, $\mu$ and $\tau$ final state leptons \cite{Achard:2001qw}. While bounds on stable neutrinos are set by the Z invisible width at half of the Z mass \cite{pdg}\cite{Lenz:2010ha}. The existence of mixed mass terms in the fourth generation lepton sector however,  allows for even lighter neutrino masses, the bounds are as low as 62.1 GeV for unstable neutrinos which decay to $\tau$ leptons \cite{Carpenter:2010dt}, and below half of the Z mass, 33.5 for stable neutrinos \cite{Carpenter:2010sm}.  Many collider searches for fourth generation leptons have been proposed with a variety of interesting final state topologies such as jets plus missing energy, or like sign dileptons plus multi-jets.  \cite{Carpenter:2010sm}\cite{Rajaraman:2010ua}\cite{Rajaraman:2010wk}\cite{Carpenter:2010bs}.  In addition, the fact that fourth generation neutrinos may be fairly light offers an exciting possibility for Higgs physics, that a Higgs with mass under 160 GeV may dominantly decay to these states if they are kinematically accessible. Such a possibility was explored both for unstable neutrinos and for stable neutrinos by \cite{Keung:2011zc}\cite{Whiteson:2011cm} \cite{Belotsky:2002ym} as well as \cite{Cetin:2011fp}.

Current Higgs searches seem to severely constrain fourth generation scenarios.  At the two signa level, ATLAS rules out the standard model Higgs with masses between 155-190 GeV  and 300-450 \cite{Crammer}. The current CMS searches rule out the Standard Model Higgs in the regions between 149 - 206 GeV and 300 - 400 GeV, and this search  is interpreted in fourth generation scenarios as ruling out a very large span of Higgs masses, between 120 and 600 GeV \cite{Koryton}.  This is because when a fourth generation is present, the Higgs production cross section is enhanced by a large factor, of almost 9 over the Standard Model, due to the loop contributions of fourth generation quarks \cite{Kribs:2007nz}\cite{Li:2010fu}. If the Higgs then decayed abundantly into the standard $WW^{*}$ channel which the CMS search relies heavily on, as it does in some SM4 models, the scenario would be highly constrained by the LHC Higgs searches.  This situation is highly constraining to fourth generation models. The situation may be ameliorated by extending fourth generation models to either suppress Higgs production, or soak up the Higgs branching fraction \cite{He:2011ti}, however apparently excluded regions of standard fourth generation parameter may be shown viable without the addition of extra fields.

This paper addresses the possibility that in a fourth generation scenario the Higgs may decay dominantly to fourth generation neutrinos which are less than half the Higgs mass.  Therefore a Higgs with mass under 160 GeV may effectively decay into hidden channels, and would not have been excluded by current Higgs searches.  Diverging from \cite{Cetin:2011fp} this paper considers the constraints and presents exclusions on the general Fourth Generation lepton parameter space.  This work details the allowed fourth generation parameter space still open, and finds that Higgs masses under 155 GeV are allowed. It is found however, that the allowed parameter space is highly sensitive the neutrino mixing angle, as the Higgs coupling to fourth generation neutrinos is quite dependent on their Majorana/ Dirac field content.  This paper proceeds as follows, Section 2 reviews fourth generation neutrino masses and couplings, Section 3 discussed Higgs production and decay, Section 4 details the constrains on fourth generation parameter space from Higgs searches and Section 5 concludes.

\section{Fourth Generation Neutrino Masses and Couplings}

In the most general scenario fourth generation neutrinos have both Dirac and Majorana masses.   Using the notation of \cite{Katsuki:1994as}, the Lagrangian for neutrinos is

\bea
L =  m_D\overline{L_4}N_R +M_N N^2+\frac{m_D}{v}H\overline{L_i}N_R
\eea

The neutrino mass matrix is given by,
\begin{eqnarray}
{\cal L}_m=-{1\over 2}\overline{(Q_R^c
N_R^c)}\left(\begin{array}{cc}0&m_D\\m_d&M\end{array}\right)
\left(\begin{array}{c}Q_R\\ N_R\end{array}\right)+h.c.
\end{eqnarray}
where $\psi^c=-i\gamma^2\psi^*$.
Diagonalizing the mass matrix there are two Majorana neutrinos with different mass eigenvalues:
\bea
 \nonumber M_1=-(M/2)+ \sqrt{m_D^2+{M^2/4}} \\ M_2=(M/2)+ \sqrt{m_D^2+{M^2/4}}
\eea

The mass eigenstates can be expressed in terms of the left-right eigenstates as

 \bea
 N_1=\cos\theta Q_L^c+\sin\theta N_R+\cos\theta  Q_L+\sin\theta N_R^c
 \\
 N_2=-i\sin\theta Q_L^c+i\cos\theta  N_R+i\sin\theta Q_L-i\cos\theta  N_R^c
  \eea

\noindent
  with neutrino mixing angle
\bea
\nonumber \tan\theta=m_1/m_D
\eea

The Higgs couples to the neutrino mass eigenstates  proportional to powers of the neutrino mixing angle.
The Higgs coupling to neutrinos pairs is given by
\bea
\frac{m_D}{v_h}H N_1 N_1 sin(2\theta)+\frac{m_D}{v_h}H N_2 N_2 sin(2\theta)+\frac{m_D}{v_h}H N_1 N_2 2i cos(2\theta).
\eea
The neutrino mixing angle varies between $\pi / 4$ for pure Dirac-type
neutrinos and $\pi/2$ for Majorana-type. We see that in the limit
where the neutrinos are pure Majorana, $m_D$ approaches zero, and the Higgs decouples from neutrinos.   While for pure Dirac states the coupling is maximal.

The neutrinos couple to the charged and neutral currents like $gW_\mu^+ J^{\mu +}
 +gW_\mu^- J^{\mu -}+gZ_\mu J^{\mu}$ where
\bea
 J^{\mu }
= {1\over 2\cos\theta_W}(-c^2_\theta\bar N_1\gamma^\mu \gamma^5N_1
-2is_\theta c_\theta\bar N_1\gamma^\mu N_2-s^2_\theta\bar N_2 \gamma^\mu\gamma^5N_2))
\\
J^{\mu +}=
c_i\overline{(c_\theta N_1-i s_\theta N_2)}\gamma^\mu l^i_L~~~~~~~~~~~~~~~~~~~
\eea

The heavy neutrino may decay to the light neutrino through an or offshell Z. The lightest neutrino may be absolutely stable, or it may decay to a standard model charged lepton though an on or offshell W.

\section{Higgs production and decay}

The greatest source of Higgs production at hadron colliders is a single Higgs produced by gluon fusion through a heavy quark loop.  The $gg\rightarrow h$ production cross section is given by

\bea
\label{production}
\sigma (g g  \to h) &=& \frac{1}{16}\Gamma (h\to gg) \frac{16 \pi^2}{s m_h} \>\int_{m_h^2/s}^1 \frac{dx}{x} g(x) g(\frac{m_h^2}{s  x}),
\eea

where $\Gamma (h\to gg)$ is the Higgs to gluon decay width

\bea
 \Gamma(h\rightarrow gg)=\frac{\alpha_s G_F m_h^3}{16 \sqrt{2} \pi^3}\Sigma_i( {\tau_i}(1+(1-\tau_i)f(\tau_i)))
\eea
with
\bea
\tau_i=\frac{4 m_{f_i}^2}{m_h^2},  f(\tau_i)= ({sin}^{-1}\sqrt{1/\tau_i})^2
\eea
Here, one is summing over all heavy flavors of quarks.  In fourth generation models, there exist both an extra heavy up and down type quark with large Yukawa couplings which contribute to the Higgs gluon decay width. This decay width is thus substantially increased over that of the Standard Model by a factor larger than 8.  For calculations of Higgs production cross sections in the SM see for example \cite{Baglio:2010ae} and \cite{Li:2010fu} in fourth generation models.  The Higgs production cross section at LHC is quite large, remaining over 10 picobarns for Higgs masses up to $\sim 500$ GeV.

The main LHC Higgs decay channels in which one looks for a Higgs produced though gluon fusion are $h\rightarrow \gamma \gamma$ and  $h\rightarrow WW$.  The Higgs decay width into $WW^{*}$ are given in \cite{Rizzo:1980gz}
and are proportional to

\bea
 \Gamma(h\rightarrow VV)\sim \frac{ 3 {G_F}^2 m_V^4 m_h}{16 \pi^2}
\eea

These begin And begins to take up a large portion of the Higgs branching fraction for Higgs masses just under twice the W mass.  Once the vector bosons are on shell, they completely dominate the Higgs branching fraction.

\begin{figure}[h]
\centerline{\includegraphics[width=8 cm]{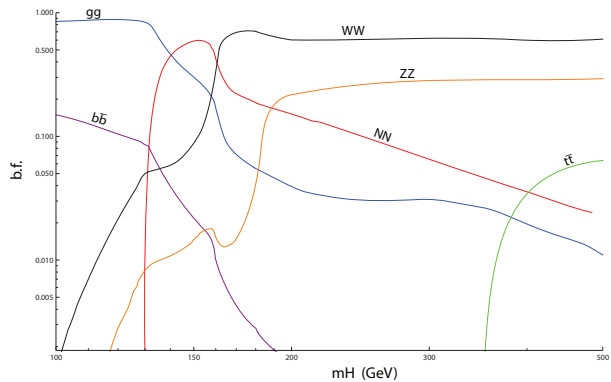}}
\caption{Higgs branching fractions vs Higgs mass in a fourth generation scenario for the benchmark point $m_{n_1} = 63 GeV$, $m_{n_2}=300 GeV$. }
\label{fig:2pgm}
\end{figure}

The Higgs may decay to photons through a one loop process involving with W bosons and charged fermions the loop.   It has been noted that Higgs branching fraction to photons in fourth generation models is suppressed relative to the standard model due to destructive interference between gauge bosons and heavy quarks in the loop.  Pervious work has claimed that the total Higgs branching fraction to photons is suppressed relative to the SM by a factor of 9 \cite{Kribs:2007nz}.  In fourth generation scenario is, as discussed above, the effective Higgs coupling to gluons is substantially increased over the standard model.  This means that the Higgs decay channel to gluons can take up a large portion of the branching of a light Higgs.

Higgs decays to fermions are proportional to their Yukawa couplings and are given by

 \bea
\Gamma_D = \frac{m_h y_D^2}{8 \pi}(1-\frac{4m_d^2}{m_h^2})^{3/2}
\eea

In the standard model, the dominant fermionic decay model of the Higgs is to the third generation quarks.  If the top is kinematically inaccessible, the largest fermionic branching fraction will be to bottoms.  However, the bottoms do not have a particularly large Yukawa coupling, if other heavy states are kinematically accessible they may easily overtake to bottom branching fraction. As seen in Section 2, the Higgs couples to fourth generation neutrinos. This coupling varies depending on neutrino mixing angle, and is maximized for pure Dirac type neutrinos.  Stable or unstable neutrinos may be quite light, and if the neutrinos are under half of the Higgs mass, the Higgs will have a large branching fraction into heavy neutrinos that have an appreciable Dirac component.  The Higgs decay width into neutrinos is given by

\bea
\Gamma_D = \frac{m_h {m_N}^2}{8 v^2 \pi}sin{2 \theta}(1-\frac{4m_d^2}{m_h^2})^{3/2}
\eea

Figure 1 is a plot of Higgs branching fractions reproduced from Figure 2 of reference \cite{Whiteson:2011cm} in a fourth generation scenario for a benchmark point in which the lightest fourth generation neutrino is kinematically accessible for Higgs decay.  One sees that for Higgs masses under half of the W mass, fourth generation neutrinos may dominate the branching fraction.

\begin{figure}[h]
\centerline{\includegraphics[width=10 cm]{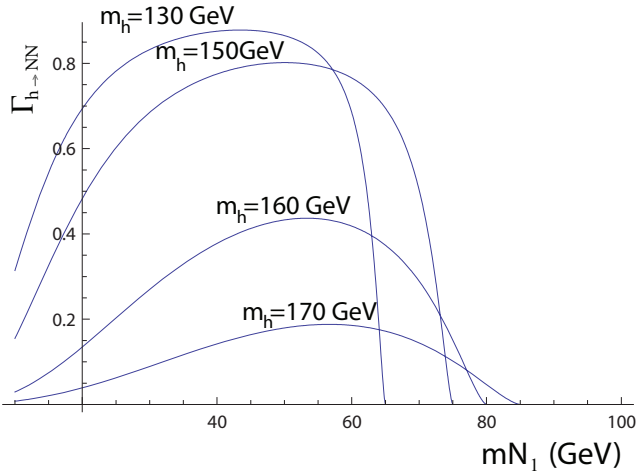}}
\caption{{Plot of the Higgs branching fraction to Fourth Generation neutrinos as a function of the mass of the lightest neutrino $N_1$.  Here The mass of the heavier neutrino $N_2$ has been set at 400 GeV.  The branching fraction is shown for Higgs mass values of 170,160,150 and 130 GeV}}
\label{fig:2pgm}
\end{figure}

\section{Higgs Mass Bounds}

The current Higgs mass bounds in fourth generation scenarios are inferred from the CMS combined search.  One can see by looking at Plot 6 of reference \cite{Koryton}, that the main Higgs exclusions tracks the $h\rightarrow WW$ decay channel fairly closely, except for a small region of parameter space around Higgs masses of 120 GeV where the Higgs to photon search has sensitivity.  However in fourth generation scenarios we have seen that the Higgs to photon branching fraction is suppressed thus we expect that only the $h\rightarrow WW$ channel can be used to extract Higgs mass constraints in fourth generation scenarios.

Current Higgs mass exclusions are claimed in fourth generation scenarios from 120-600 GeV.  These exclusions are much larger than those claimed for the Standard Model due to the fact that fourth generation scenarios are assumed to produce many more $h\rightarrow WW^{*}$ events than the Standard Model.  The total number of such events - proportional to $\sigma_{gg\rightarrow h_{SM4}} \times \Gamma_{h\rightarrow {WW^{*}}_{SM4}}$- is much larger than the number expected in the Standard Model due to the extremely large Higgs production cross section.


However, the Higgs mass exclusion for the fourth generation was made under the assumption that the there were no additional light states that the Higgs could decay into.  We have seen that fourth generation neutrinos can make up the dominant branching fraction for a Higgs with mass under twice $m_W$.   Figure 2 shows the Higgs branching fraction vs. light neutrino mass for a variety of Higgs masses.  One see that the Higgs branching fraction into neutrinos is very high for light Higgs masses, and doesn't drop substantially until the onshell gauge bosons become kinematically accessible to the Higgs.  Standard Higgs searches are not sensitive to Higgs final states which result from neutrino decay; if the Higgs decays to the stable lightest neutrinos, its decay will be invisible, in the case of unstable neutrinos, the Higgs will decay via the process $h\rightarrow N N \rightarrow \ell \ell W^{*}W^{*}$, a decay chain most likely ending in hard leptons plus multi-jets.

One may see that the Higgs branching fraction to offshell gauge bosons is reduced by a substantial factor in Fourth Generation scenarios with light neutrinos as opposed to scenarios without them.  Figure 3a shows the ratio of Higgs to $WW^{*}$ branching fractions of two Fourth Generation scenarios, one with neutrinos kinematically accessible for Higgs decays and the other entirely without them. The ratio of branching fractions is everywhere less than 1, and is quite small in much of the parameter space.  One can see the Higgs mass exclusion down to 120 GeV only holds where the Higgs has effectively {\it no} decay width into onshell fourth generation neutrinos.

Alternatively, Figure 3b shows the ratio of Higgs branching fractions to $WW^{*}$ in the Fourth Generation with light neutrinos to the Standard Model.  Again, in a large region of parameter space, this ratio may be quite small.  In fact, in some region of parameter space, the branching fraction of Higgses to offshell gauge bosons may be reduced by more than the factor of 8-9 by which the Higgs production cross section was increased by adding a fourth generation of particles.  In this way the total number of $h \rightarrow WW^{*}$ events from the Fourth Generation need not be extremely increased over the number of events one expects from the Standard Model.  Thus the claim of Higgs exclusions down to 120 GeV does not apply.

To see what regions of Fourth Generation space are ruled out by the current Higgs search, we must calculate the total number of $h\rightarrow WW^{*}$ events for each point in parameter space and compare to the CMS exclusion limits for the $h \rightarrow WW{*}$ search.  One sees that in some regions of parameter space, the $h\rightarrow WW^{*}$ branching fraction is quite small. However, the Higgs coupling to neutrinos (and hence the relevant branching fraction) is also very sensitive to the neutrino mixing angle, and is different at each point in the neutrino mass plane.

\begin{figure}[h]
\begin{center}$
\begin{array}{cc}
\includegraphics[width=7.8 cm]{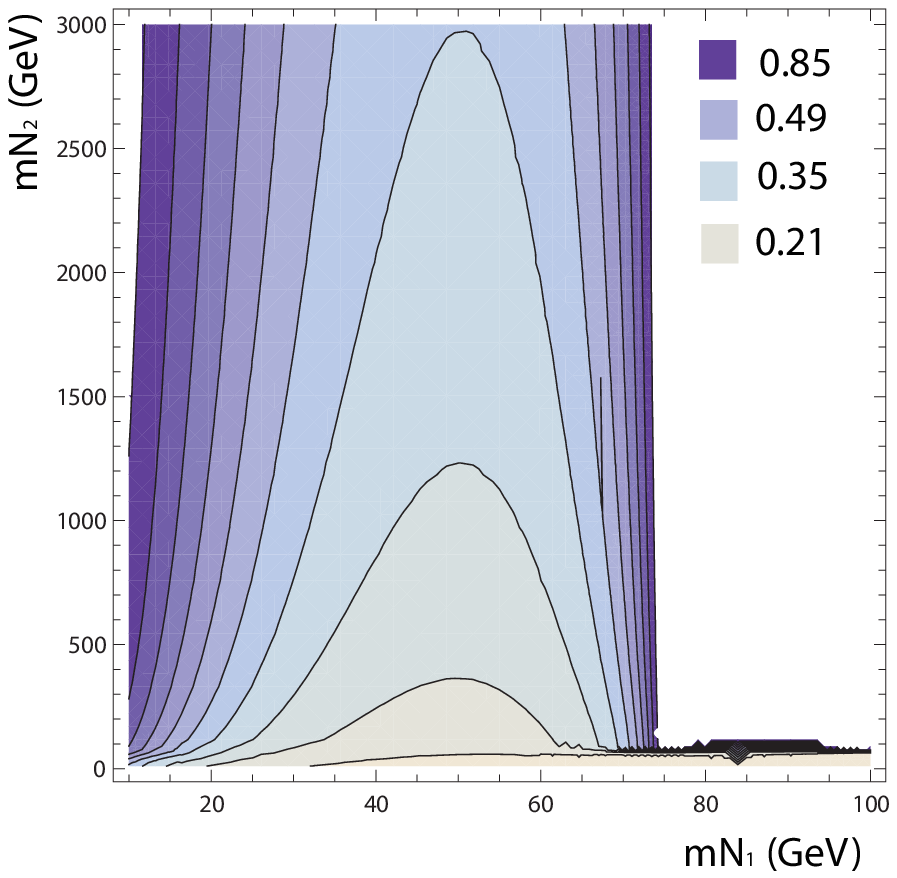} &
\includegraphics[width=10.0 cm]{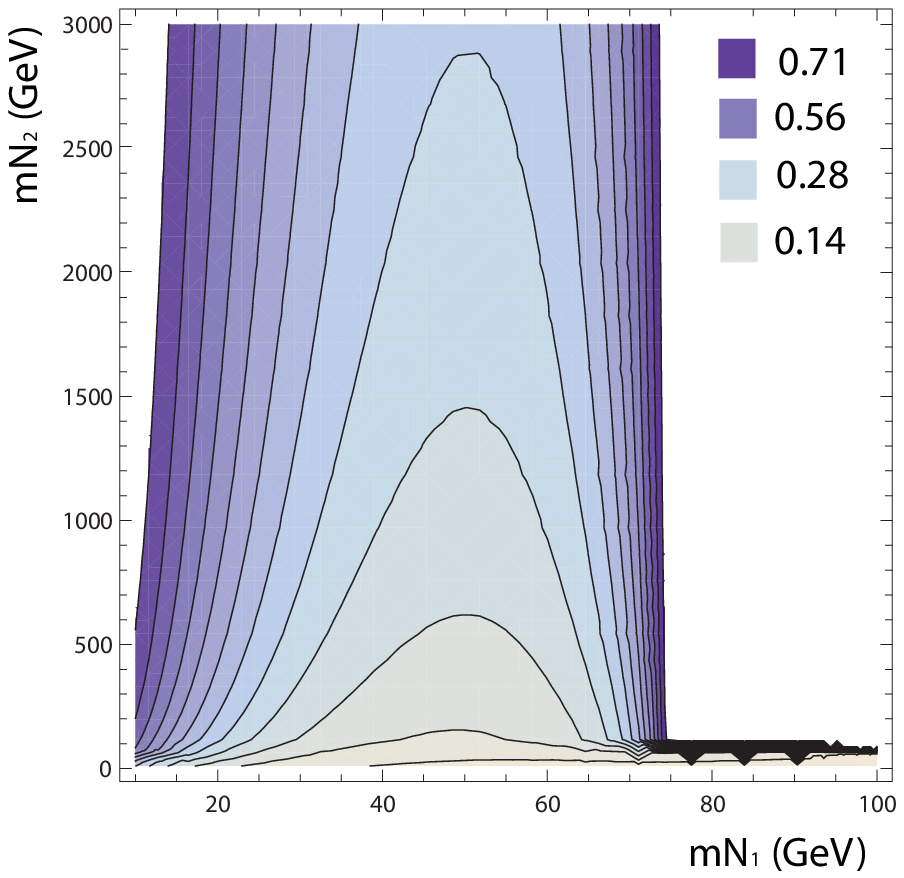}
\end{array}$
\end{center}
\caption{ Contour plot in the $N_1 N_2$ mass plane. Right: ratio between the decay widths $h\rightarrow WW^{*}$ in a fourth generation model with Majorana neutrinos light vs. in the Standard Model.  Left: ratio between the decay widths $h\rightarrow WW^{*}$ in a fourth generation model with Majorana neutrinos light vs. in a fourth generation model without light Majorana neutrinos.  Plots are shown for a benchmark Higgs mass of 150 GeV Higgs}
\end{figure}

Figure 5 shows the allowed regions of parameter space in the neutrino mass plane for several contours of Higgs mass values.  Of course no parameter space is present for neutrinos greater than half of the Higgs mass.  There is a sharp falloff in the amount of allowed parameter space once the Higgs becomes sufficiently massive, here the gauge bosons are getting closer to being onshell and beginning to take up more of the Higgs branching fraction.  Notice that there is parameter space for Higgs up to 155 GeV.  Also, the most parameter space  allowed is greatest where the neutrinos are the most well mixed between Dirac and Majorana masses.  As the neutrinos become more Majorana, the mass difference between the neutrino states grows and the coupling to the Higgs decreases, decreasing the Higgs branching fraction into neutrinos.

\begin{figure}[h]
\centerline{\includegraphics[width=10 cm]{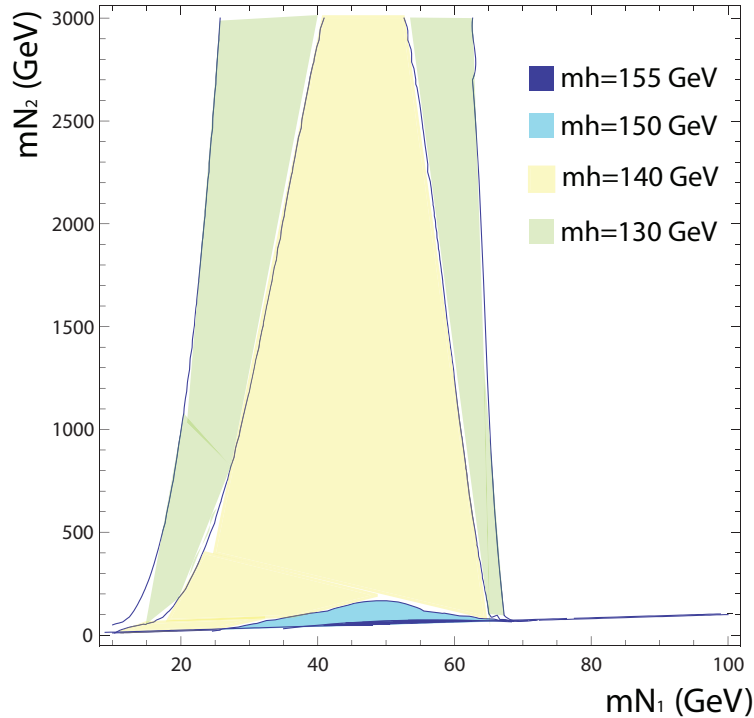}}
\caption{{Allowed regions in n1 n2 mass plane.}}
\label{fig:2pgm}
\end{figure}

The Higgs branching fraction at high Higgs mass is almost entirely taken up by onshell diboson production.  One can see that the Higgs decay width to dibosons goes like $mh^3$.  As the Higgs becomes much heavier than twice the gauge boson mass, the branching ratio basically becomes a constant. In this regime, the branching ratio to heavy fermions can not overtake the diboson ratio.  As one increases the Yukawa coupling, which is proportional to the dirac neutrino mass component, one increases the mass of the neutrino and will pay a phase space factor as a consequence.  Thus at larger Higgs masses, the Higgs branching fraction will be overwhelmingly taken up by the gauge bosons. Only a few percent of the branching ratio is taken up by fourth generation neutrinos at high Higgs mass.  Therefore one expects that the upper bound of the Higgs exclusion at 600 GeV must hold.

\section{Conclusions}

There exists fourth generation parameter space in which a Higgs may dominantly decay to fourth generation neutrinos. For Higgs masses under 155 GeV current LHC searches do not rule out fourth generation scenarios despite that fact the Higgs production in such a scenario is increased over standard model production by a substantial factor.  The Higgs may decay invisibly if fourth generation neutrinos are stable or, if the neutrinos decay to W bosons and standard model leptons, the Higgs may decay to dibosons plus hard leptons.

Finding a Higgs in such a scenario could be quite challenging and presents an interesting avenue for further work.  The authors of \cite{Keung:2011zc} have inquired into the situation of looking at Higgs decays  in the case of stable lightest neutrino. This however required the Higgs to decay into the second lightest neutrino which then cascaded $h\rightarrow N_2 N_2 \rightarrow ZZ N_1 N_1$.  If the second neutrino is not kinematically accessible, the Higgs signal will be pure missing energy.  In the case that the lightest neutrinos are unstable, the Higgs may decay to dibosons plus hard leptons $h\rightarrow N_1 N_1 \rightarrow WW \ell \ell$.  If the Higgs is under 160 GeV, existing inclusive like-sign dilepton searches constrain the final state lepton in this scenario to be a tau.  The case that the Higgs decays to like sign taus plus jets presents a challenging but interesting search.

\section{Acknowledgement}
The work was supported by NSF grant PHY-0709742.


\end{document}